**Carbonate-Bridged Dinuclear and Trinuclear Dysprosium(III) Single-Molecule Magnets**

Guang Lu, Yan-Cong Chen, Si-Guo Wu, Guo-Zhang Huang, Jun-Liang Liu, Zhao-Ping Ni,* and Ming-Liang Tong*

MOE Key Lab of Bioinorganic and Synthetic Chemistry, School of Chemistry, Sun Yat-Sen University, Guangzhou 510275, P. R. China

**ABSTRACT:** In 2016, we reported a single-ion magnet [Dy(bbpen)Br] with an energy barrier over 1000 K. Here a dimeric [Dy$_2$(μ-CO$_3$)(bbpen)$_2$(H$_2$O)]·H$_2$O·CH$_3$OH (**1**) and a trimeric [Dy$_3$(μ$_3$-CO$_3$)(bppen)$_3$](CF$_3$SO$_3$)·H$_2$O (**2**) single-molecule magnets (SMMs) were obtained through replacing the Br$^−$ anion with the CO$_3^{2−}$ bridge. Their effective relaxation barriers at zero dc field are decrease to 51 K and 422 K, respectively, which are consist with their structural modifications.

**INTRODUCTION**

Single-molecule magnets (SMMs) with the ability of slow relaxation in the molecular level have attracted great attentions due to their potential applications in ultra-high density data storage, quantum computing and molecule spintronics.[1-7] However, high-performance SMMs in practice require high effective relaxation barrier ($U_{eff}$) and high blocking temperature ($T_B$).[8] Starting from [Mn$_{12}$O$_{12}$(OAc)$_{16}$(H$_2$O)$_4$] {Mn$_{12}$Ac} with $U_{eff}$ = 61 K and $T_B$ ~ 3 K,[2] the records were frequently refreshed.[9-12] The highest $U_{eff}$ now reaches 1837 K, which is a [Dy(Cp$^{ttt}$)$_2$]$^+$ organometallic complex.[13-14]

In order to achieve SMMs from a bottom-up approach, one effective strategy is to introduce the magnetic interactions between the performant and stable building blocks through bridging ligands. Inspired by our reported pentagonal bipyramidal Dy$^{III}$ single-ion magnet (SIM), [Dy(bbpen)Br] (H$_2$bbpen = *N*,*N*'-bis(2-hydroxybenzyl)-*N*,*N*'-bis(2-picolyl)ethylenediamine) with $U_{eff}$ > 10$^3$ K and high stabilities as well.[15] herein the Br$^−$ anion in [Dy(bbpen)Br] was replaced by the CO$_3^{2−}$ bridging ligand. Consequently, [Dy$_2$(μ-CO$_3$)(bbpen)$_2$(H$_2$O)]·H$_2$O·CH$_3$OH (**1**) and [Dy$_3$(μ$_3$-CO$_3$)(bppen)$_3$](CF$_3$SO$_3$)·H$_2$O (**2**) were successfully isolated by different synthetic methods. Magnetic dynamics are distinct for the both compounds whilst their bppen$^{2-}$ parts are similar to that in [Dy(bbpen)Br].

**RESULTS AND DISCUSSION**

**Crystal Structures**

Crystals of **1** and **2** are isolated through slow evaporation and solvothermal synthesis, respectively. Single-crystal X-ray diffraction revealed that complexes **1** and **2** crystallize in triclinic space group *P*-1 and monoclinic *P*2$_1$/*n*, respectively. In **1**, two Dy$^{III}$ ions are bridged through one CO$_3^{2-}$ anion using $\mu$-$\eta^2$:$\eta^1$-modes as shown in Figure 1, which result in the Dy···Dy distance of 6.205 Å. However, only one crystallographic independent Dy$^{III}$ ion exists in the crystal structure. The asymmetric unit of **1** consists of half of the molecule with disordered CO$_3^{2-}$ anion and water. Therefore, Such Dy$^{III}$ ion is eight-coordinate with the N$_4$O$_4$ donor set, which equally comes from one bbpen$^{2-}$ ligand, one monodentate CO$_3^{2-}$ anion and one water or from one bbpen$^{2-}$ ligand and one bidentate CO$_3^{2-}$ anion. Its geometry is close to the biaugmented trigonal prism. Importantly, the coordination mode of bbpen$^{2-}$ in **1** is distorted but similar to that in [Dy(bbpen)Br]. However, its average axial Dy−O distance is 2.229 Å, which is slightly longer than 2.163 Å in [Dy(bbpen)Br]. Meanwhile, compound **1** shows a smaller O−Dy−O bond angle of 149.26°.

In **2**, three Dy$^{III}$ ions are connected by one CO$_3^{2-}$ anion using $\mu_3$-$\eta^2$:$\eta^2$:$\eta^2$-mode (Figure 1), and then, each Dy$^{III}$ ion is surrounded by one bbpen$^{2-}$ anions. Therefore, each Dy$^{III}$ ion is coordinated by the N$_4$O$_4$ donor set with a distorted triangular dodecahedron configuration. The intramolecular Dy···Dy distances are 4.811 Å, 4.837 Å and 4.879 Å, giving a nearly perfect equilateral triangle. The coordination mode of bbpen$^{2-}$ in **2** is also distorted but similar to that in [Dy(bbpen)Br]. Their average axial Dy−O distance is 2.201 Å, 2.188 Å, and 2.195 Å for Dy1~Dy3, respectively, which are all longer than that in [Dy(bbpen)Br]. Meanwhile, their O−Dy−O bond angles are 161.06°, 162.94°, and 159.21°, respectively, which are more linear than 155.8° in [Dy(bbpen)Br].

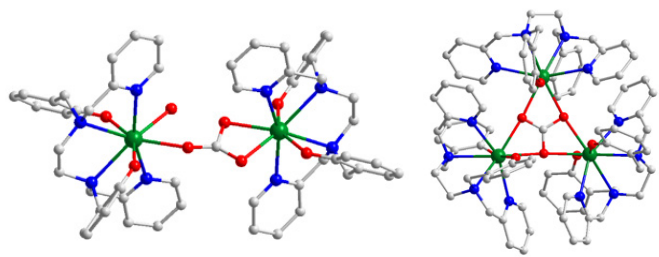

**Figure 1.** Molecular structures for **1** (left) and **2** (right). Dy, dark green; O, red; N, blue; C, grey. For Clarity, the hydrogen atoms are omitted.

**Magnetic Characterization.**

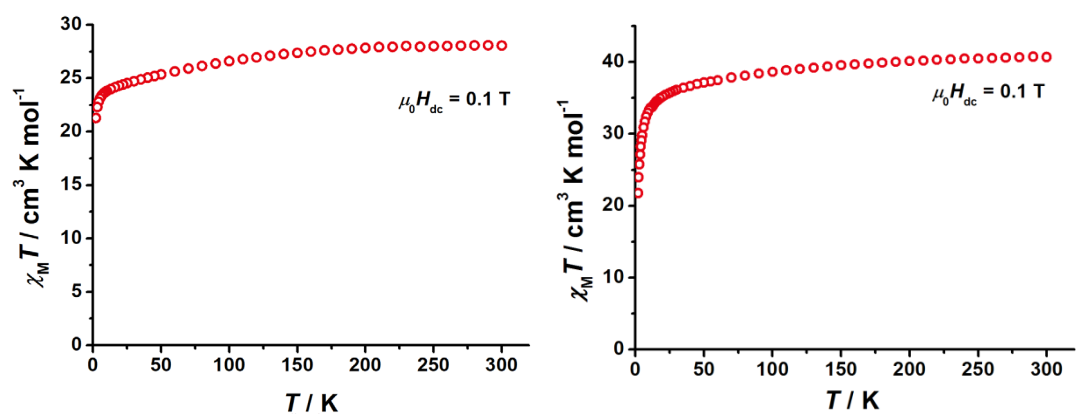

**Figure 2.** Temperature-dependent dc magnetic susceptibility for **1** (left) and **2** (right).

Temperature-dependent direct-current (dc) magnetic susceptibility measurements were performed on polycrystalline samples of **1** and **2** under 0.1 T dc field (Figure 2). The room temperature $\chi_M T$ values are 28.07 cm$^3$ K mol$^{-1}$ and 40.68 cm$^3$ K mol$^{-1}$ for **1** and **2**, respectively, which are close to the expected value of 14.17 cm$^3$ K mol$^{-1}$ per Dy$^{III}$ ion ($S = {}^5/_2$, $L = 5$, $^6H_{15/2}$, $J = {}^{15}/_2$, $g = {}^4/_3$). On cooling, both of the $\chi_M T$ products decrease slightly to 21.29 cm$^3$ K mol$^{-1}$ and 21.75 cm$^3$ K mol$^{-1}$ at 2 K, which suggests the presence of antiferromagnetic interactions and/or the strong crystal-field splitting of the $^6H_{15/2}$ term. The isothermal magnetization (Figure S1) for both complexes increases rapidly at low field and reach ~5 $N\beta$ per Dy$^{III}$, indicating strong magnetic anisotropy.

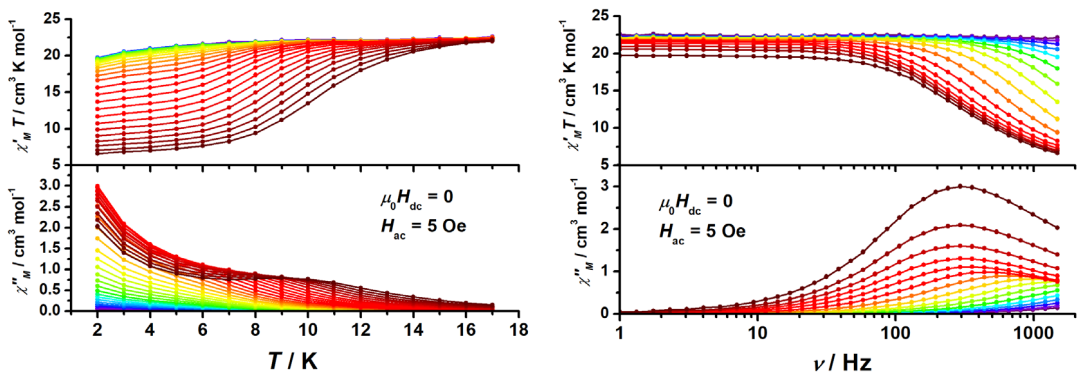

**Figure 3.** Temperature-dependent (left) and frequency-dependent (right) ac magnetic susceptibility for **1** at zero dc field. The lines are guides for the eyes.

Alternating-current (ac) magnetic susceptibilities were then measured to probe the dynamic magnetic properties. For **1** at zero dc field (Figure 3), the out-of-phase ac signals rise below ~17 K but show obvious "tails" at low temperature. Such phenomenon usually indicates the presence of fast relaxation process such as the quantum tunneling of magnetization (QTM), which is also evidenced by the frequency-dependent ac peaks.

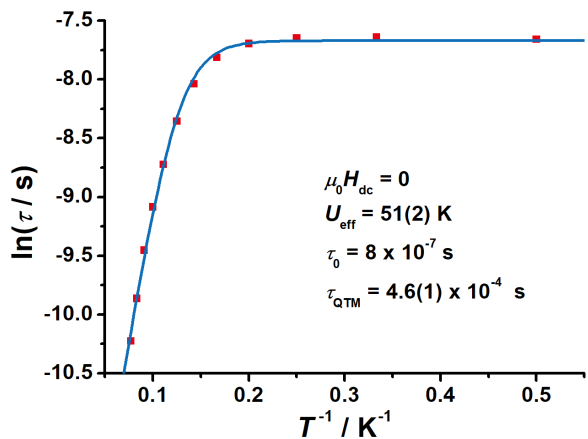

**Figure 4.** Temperature-dependent relaxation times for **1** at zero dc field.

To obtain the relaxation times ($\tau$), the ac data is fitted using the generalized Debye model. Above ~5 K, the relaxation time of **1** increase rapidly with the decrease of temperature, but it becomes almost constant at low temperature. Such common temperature dependency can be fitted with the combination of Orbach process and QTM, namely $\tau^{-1} = \tau_0^{-1}\exp(-U_{\text{eff}}/T) + \tau_{\text{QTM}}^{-1}$, where the best fit gives an effective energy barrier $U_{\text{eff}} = 51(2)$ K and a $\tau_{\text{QTM}} = 0.46(1)$ ms (Figure 4).

To suppress the fast QTM, an external dc field is applied at 2 K (Figure S2). Interestingly, the peak of out-of-phase ac susceptibility at high frequency slowly disappears, while a new peak at low frequency emerges. The field-dependent relaxation time show a peak at 0.2 T, which is selected as the optimized field for further ac magnetic characterization.

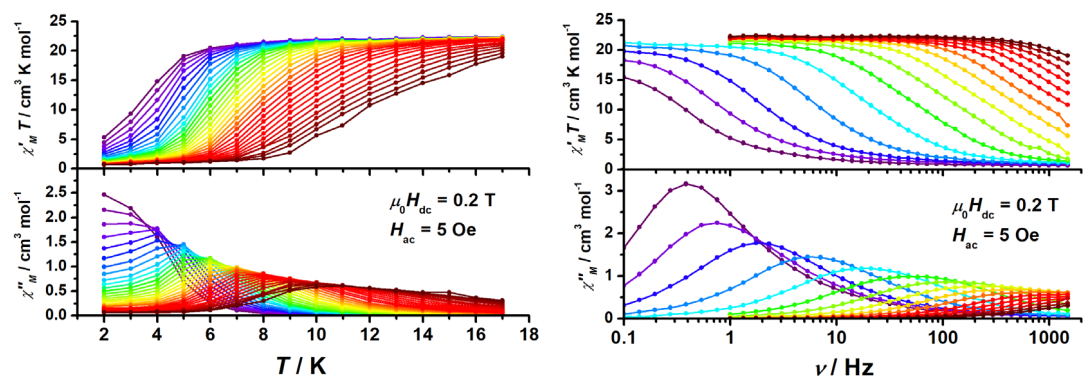

**Figure 5.** Temperature-dependent (left) and frequency-dependent (right) ac magnetic susceptibility for **1** at 0.2 T dc field. The lines are guides for the eyes.

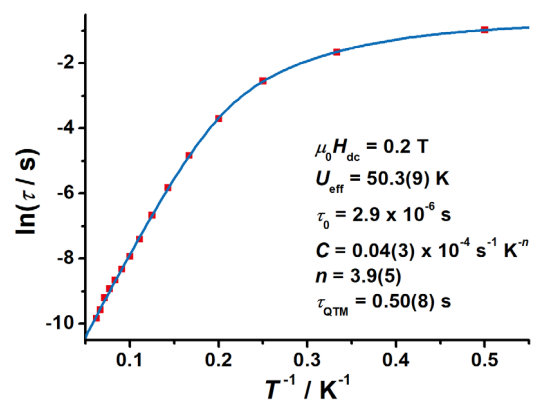

**Figure 6.** Temperature-dependent relaxation times for **1** at 0.2 T dc field.

Under a 0.2 T dc field, the QTM at low temperature is greatly suppressed (Figure 5), and the out-of-phase peak frequency moves to as low as ~0.4 Hz at 2 K. The relaxation times at low temperature are now dominated by a Raman process (Figure 6), which can be fitted with the multiple relaxation equation $\tau^{-1} = \tau_0^{-1}\exp(-U_{eff}/T) + CT^n + \tau_{QTM}^{-1}$. The best fit gives an effective energy barrier $U_{eff}$ = 50.3(9) K which is similar to that in zero field, a typical Raman exponent parameter $n$ = 3.9(5), and finally a much longer $\tau_{QTM}$ = 0.50(8) s.

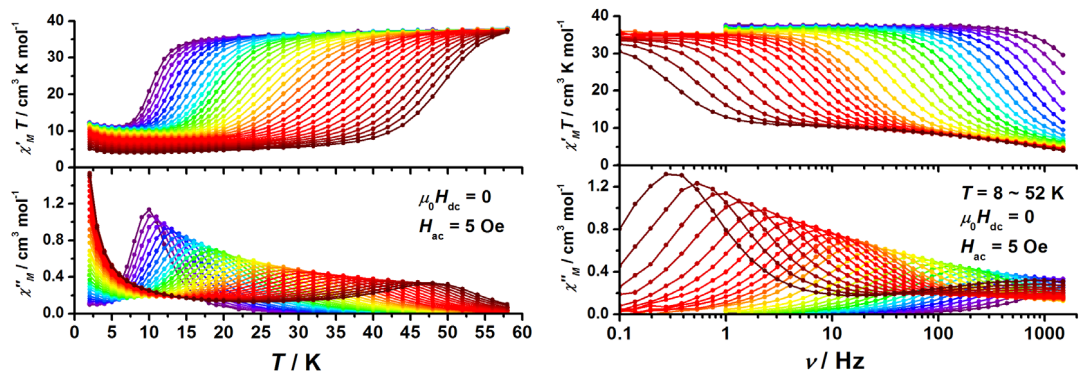

**Figure 7.** Temperature-dependent (left) and frequency-dependent (right) ac magnetic susceptibility for **2** at zero dc field. The lines are guides for the eyes.

The magnetic dynamics for **2** shows largely different behaviors. At zero dc field (Figure 7), the highest temperature-dependent peak of the out-of-phase ac susceptibility move to 48 K (1488 Hz), and that of 1 Hz is located at 10 K. From the frequency-dependent plot, it is clear that the peaks keep moving to lower frequency with the decreasing of temperature. At lowest temperatures, a secondary peak is visible at high frequency, which may due to the intramolecular Dy…Dy interactions that lead to additional excited states.

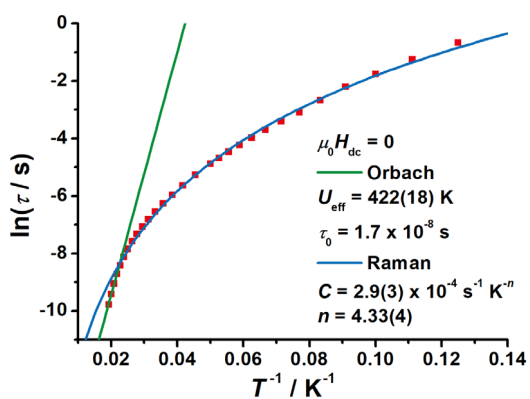

**Figure 8.** Temperature-dependent relaxation times for **2** at zero dc field.

The ac data for **2** is fitted using the generalized Debye model to obtain the relaxation times (Figure 8). The temperature dependency of relaxation times does not show any obvious sign of QTM. At high temperature, the almost linear curve indicates an Arrhenius behavior, which can be fitted as an Orbach process with $U_{eff}$ = 422(18) K. This energy barrier is much larger than that of complex **2**, but it only dominates a small fraction of temperature region. At lower temperature, the relaxation times are dominated by a Raman process with $n$ = 4.33(4), down to the low-frequency limit of ac magnetometer.

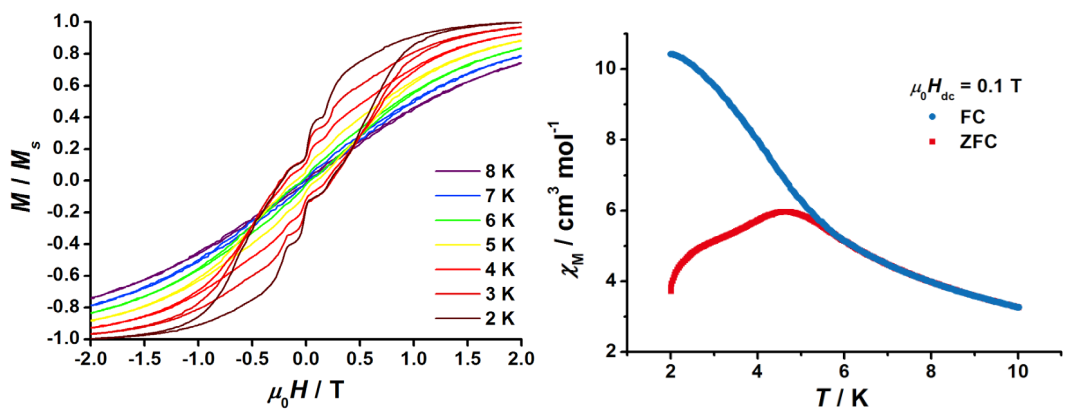

**Figure 9.** Magnetic hysteresis loops at 0.02 T/s (left) and FC-ZFC magnetic susceptibility under a 0.1 T dc field (right) for **2**.

As the relaxation times for **2** are quite long at low temperature, magnetic blocking is expected. Indeed, the opening of the magnetic hysteresis loops is observed below 7 K, and the zero-field-cooled and field-cooled (FC-ZFC) magnetic susceptibility show clear divergences below 5.5 K (Figure 9). Interestingly, there are two sets of obvious steps on the magnetic hysteresis loops: the one near zero field is mainly attributed to the single-ion behavior, and the one at ~ 0.2 T may due to the Dy…Dy interactions.

Due to the effective suppression of QTM by magnetic interactions, the ac susceptibility for **2** does not show much field dependency (Figure S3). Nevertheless, the relaxation time reaches a peak at ~0.12 T then falls to a minimum at 0.2 T, which is in good agreement with the steps on the magnetic hysteresis loops.

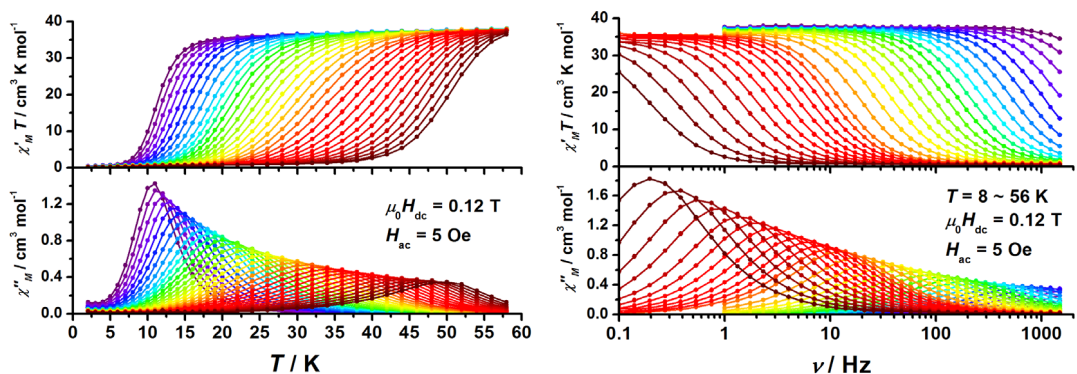

**Figure 10.** Temperature-dependent (left) and frequency-dependent (right) ac magnetic susceptibility for **2** at 0.12 T dc field. The lines are guides for the eyes.

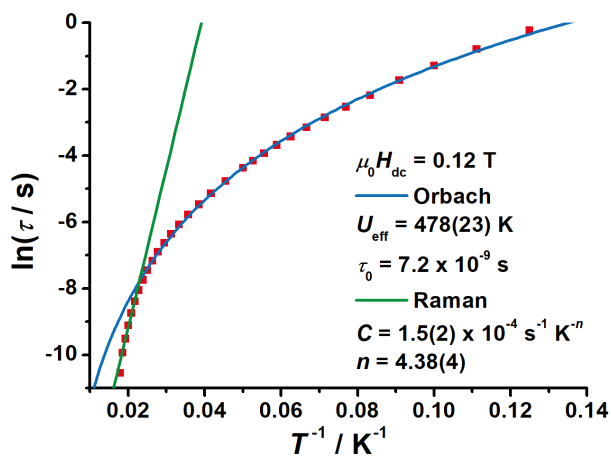

**Figure 11.** Temperature-dependent relaxation times for **2** at zero dc field.

Under a 0.12 T dc field, the ac susceptibility of **2** is generally similar to those in zero dc field, only without the secondary fast relaxation at high frequency (Figure 10). In the same way, the ac data is fitted using the generalized Debye model and the temperature dependency of relaxation times shows similar behavior (Figure 11): an Orbach process with $U_{eff}$ = 478(23) K and a Raman process with $n$ = 4.38(4). We notice that the energy barrier and Raman exponent parameter are basically consistent with those in zero field, but the Raman coefficient $C$ become much smaller and result in longer relaxation time at low temperature.

**Magnetostructural Correlations**

At zero dc field, the $U_{eff}$ greatly decreases from 1025 K for [Dy(bbpen)Br] to 51 K for **1** and 422 K for **2**. Obviously, the effective energy barrier can be greatly affected by the structural modification. Firstly, the compressed pentagonal bipyramidal $Dy^{III}$ coordination geometry is believed to form high-performance SMMs. However, **1** and **2** are eight-coordinate and they adopt distorted biaugmented trigonal prism and triangular dodecahedron configurations, respectively, which could be responsible for the main point of the decrease of effective energy barriers. Moreover, **1** has the longest Dy−O distance and smallest O−Dy−O bond angle among three compounds, where both factors are against the stability of oblate electron density for $Dy^{III}$ ion with the pure $|m_J|$ = 15/2 doublets and tend to decrease the $U_{eff}$. In the case of **2**, the factor of the large O−Dy−O bond angle would compete with the influence of the long Dy−O distance, which prevents the collapse of energy barrier like **1**.

**Conclusions**

Carbonate-bridged [Dy$_2$(μ-CO$_3$)(bbpen)$_2$(H$_2$O)]·H$_2$O·CH$_3$OH (**1**) and [Dy$_3$(μ$_3$-CO$_3$)(bppen)$_3$](CF$_3$SO$_3$)·H$_2$O (**2**) were synthesized based on the performant and stable {Dy(bbpen)} building blocks. The magnetization dynamics are very different from each other, giving the effective energy barriers of 51 K and 422 K, respectively. Most importantly, the antiferromagnetic interaction through carbonate will provide **2** to be a perfect example of single-molecule toroics (SMTs), in which the vortex-spin chirality will show in the triangular {Dy$_3$} cluster. Ab initio calculation and high-frequency/-field electron paramagnetic resonance (HF-EPR) are ongoing.

## EXPERIMENTAL SECTION

**General Procedure**. Metal salts and other reagents were commercially available and used as received without further purification. All reactions described below were performed under aerobic conditions. The H$_2$bbpen ligand (*N*,*N*′-bis(2-hydroxybenzyl)-*N*,*N*′-bis(2-picolyl)ethylenediamine) was prepared according to the reported methods. The C, H, N and S microanalyses were carried out with an Elementar Vario-EL CHNS elemental analyzer. The FT-IR spectra were recorded from KBr pellets in the range 4000−400 cm$^{-1}$ on an EQUINOX 55 spectrometer. Thermogravimetric analysis was carried out on a NETZSCH TG209F3 thermogravimetric analyzer. X-ray powder diffraction intensities for polycrystalline samples were measured at room temperature on Bruker D8 Advance Diffratometer (Cu-Kα, $\lambda$ = 1.54178 Å).

**[Dy$_2$(μ-CO$_3$)(bbpen)$_2$(H$_2$O)]·H$_2$O·CH$_3$OH (1).** A solution of DyCl$_3$·6H$_2$O (18 mg, 0.05 mmol), H$_2$bbpen (23 mg, 0.05 mmol) and triethylamine (10 mg, 0.1 mmol) in methanol (5 mL) was added into 5 mL aqueous solution of AgNO$_3$ (8 mg, 0.05 mmol), NH$_3$·H$_2$O (0.08 mL), and isonicotinic acid (6 mg, 0.05 mmol). After stirring for 2 hours, the resulting mixture was filtered, and the filtrate was left at room temperature for slow evaporation. Colourless block crystals were obtained after 1 day (yield *ca.* 10 mg, 32%).

**[Dy$_3$(μ$_3$-CO$_3$)(bbpen)$_3$](CF$_3$SO$_3$)·H$_2$O (2).** A solution of Dy(CF$_3$SO$_3$)$_3$ (30 mg, 0.05 mmol), H$_2$bbpen (23 mg, 0.05 mmol), tetramethylammonium bicarbonate (7 mg, 0.05 mmol) and triethylamine (30 mg, 0.3 mmol) in acetonitrile (9 mL) was sealed in a 23 mL Teflon-lined stainless container and heated at 75 °C for 1 day and then cooled to ambient temperature at a rate of 10 °C/h to form colourless block crystals (yield *ca.* 12 mg, 35%).

**X-ray Crystallography**. Diffraction intensities were collected on a Bruker D8 QUEST diffractometer using Mo-Kα radiation ($\lambda$ = 0.71073 Å) for **1** and **2** at 120(2) K. The structures were solved by direct methods, and all non-hydrogen atoms were refined anisotropically by least-squares on $F^2$ using the SHELXTL program suite. Anisotropic thermal parameters were assigned to all non-hydrogen atoms. Hydrogen atoms on organic ligands were generated by the riding mode.

**Magnetic Measurements**. Magnetic susceptibility measurements were collected using a Quantum Design MPMS-XL7 SQUID magnetometer and a Quantum Design PPMS-XL9 VSM. Polycrystalline samples were embedded in vaseline to prevent torqueing. AC magnetic susceptibility data measurements were performed with a 5 Oe switching field at frequencies between 0.1 and 1488 Hz. All data were corrected for the diamagnetic contribution calculated using the Pascal constants.

Corresponding Authors

*nizhp@mail.sysu.edu.cn

*tongml@mail.sysu.edu.cn

**ACKNOWLEDGEMENTS**

This work was supported by the NSFC (Grant no. 21620102002, 21771200, 21773316, 21371183, 91422302 and 21701198), the Fundamental Research Funds for the Central Universities (Grant 17lgjc13 and 17lgpy81) and the China Postdoctoral Science Foundation (National Postdoctoral Program for Innovative Talents, Grant BX201700295).

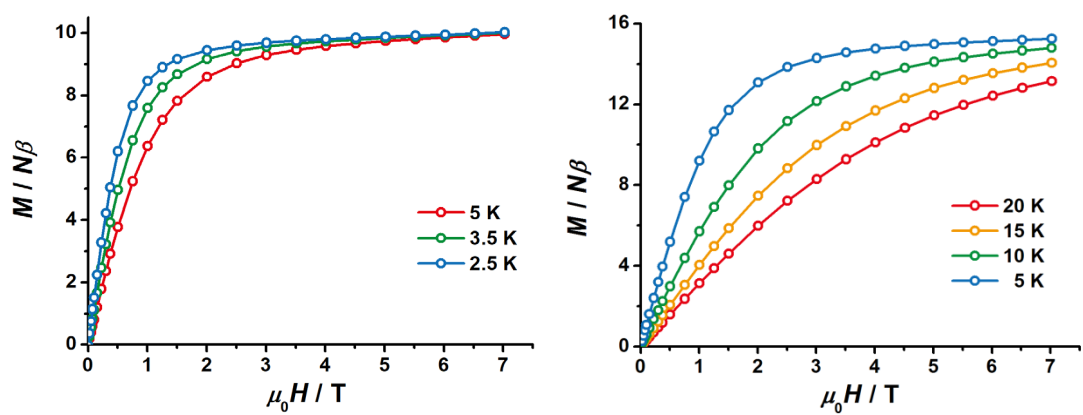

**Figure S1.** Isothermal magnetization for **1** (left) and **2** (right). The lines are guides for the eyes.

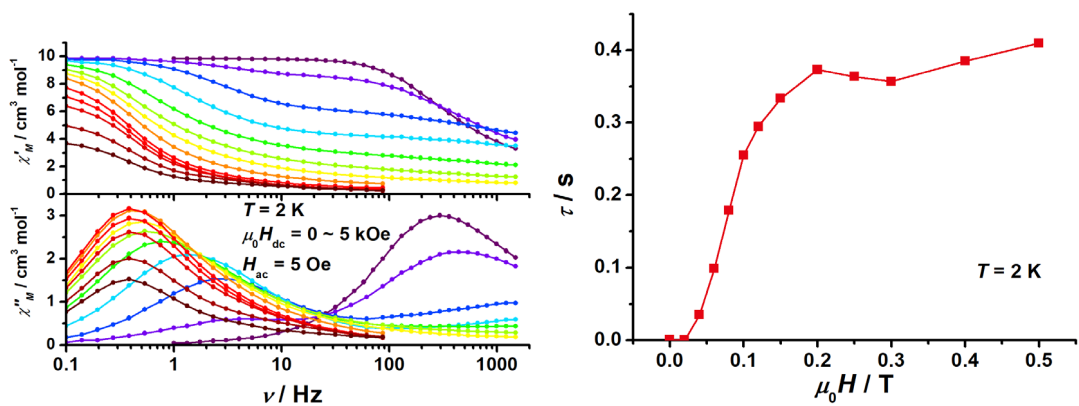

**Figure S2.** Field-dependent ac magnetic susceptibility (left) and relaxation times (right) for **1**. The lines are guides for the eyes.

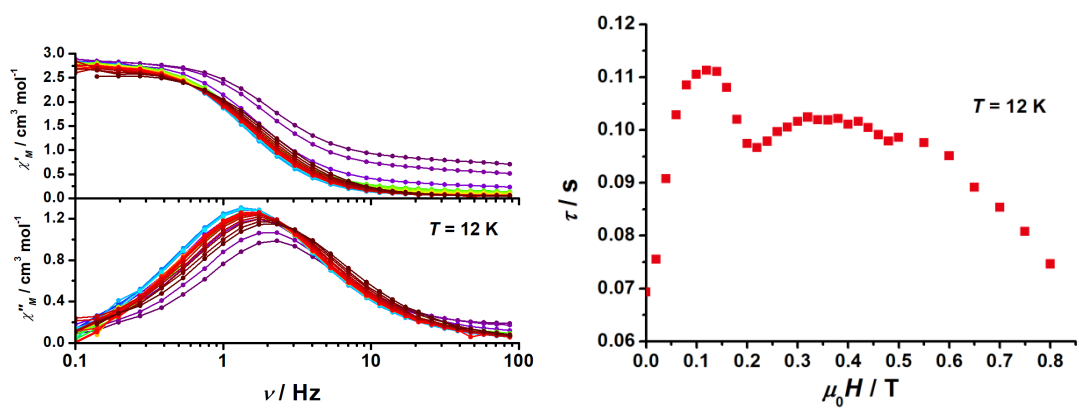

**Figure S3.** Field-dependent ac magnetic susceptibility (left) and relaxation times (right) for **2**. The lines are guides for the eyes.